\documentclass[sigconf]{acmart}
\usepackage{graphicx} 
\usepackage{balance}
\usepackage{soul} 
\usepackage{pifont}
\usepackage{makecell}
\usepackage[table,dvipsnames]{xcolor}
\usepackage{acronym} 
\newacro{IoT}{Internet-of-Things}
\newacro{RAM}{Random Access Memory}

\newcommand{\systName}{{TinyContainer}}

\copyrightyear{2026}
\acmYear{2026}
\setcopyright{cc}
\setcctype{by}
\acmConference[WiSec '26]{Proceedings of the 19th ACM Conference on Security and Privacy in Wireless and Mobile Networks}{June 30-July 03, 2026}{Saarbrücken, Germany}
\acmBooktitle{Proceedings of the 19th ACM Conference on Security and Privacy in Wireless and Mobile Networks (WiSec '26), June 30-July 03, 2026, Saarbrücken, Germany}
\acmDOI{10.1145/3765613.3811689}
\acmISBN{979-8-4007-2201-1/2026/06}

\newcommand{\todoBB}[2][]{}
\newcommand{\todoCG}[2][]{}
\newcommand{\todoSB}[2][]{}
\newcommand{\todoEB}[2][]{}

\begin{document}

\title{\systName: Container Runtime Middleware Enabling Multi-tenant Microcontrollers with Built-in Security}

\author{Bastien Buil}
\orcid{0009-0007-0062-2511}
\affiliation{%
  \institution{Orange Research}
  \city{Caen}
  \country{France}
}
\affiliation{%
  \institution{Cnam}
  \department{CEDRIC}
  \city{Paris}
  \country{France}
}
\email{bastien.buil@orange.com}

\author{Chrystel Gaber}
\orcid{0000-0001-5768-7665}
\affiliation{%
  \institution{Orange Research}
  \city{Caen}
  \country{France}
}
\email{chrystel.gaber@orange.com}

\author{Samuel Legouix}
\orcid{0009-0001-8804-2146}
\affiliation{%
  \institution{Orange Research}
  \city{Caen}
  \country{France}
}
\email{samuel.legouix@orange.com}

\author{Emmanuel Baccelli}
\orcid{0000-0001-6239-9983}
\affiliation{%
  \institution{Inria}
  \department{TRiBE}
  \city{Palaiseau}
  \country{France}
}
\email{emmanuel.baccelli@inria.fr}

\author{Samia Bouzefrane}
\orcid{0000-0002-0979-1289}
\affiliation{%
  \institution{Cnam}
  \department{CEDRIC}
  \city{Paris}
  \country{France}
}
\email{samia.bouzefrane@lecnam.net}

\begin{abstract}
Software containerization technologies for resource-limited devices enable multi-tenant microcontrollers, which allow running multiple applications with different permission levels.
However, current solutions lack run time configuration over various settings on container scheduling and container permissions to host resources. 
This limits the applicability of constrained containerization in dynamic and heterogeneous environments. This paper introduces \systName{}, a lightweight software container management middleware designed for multi-tenant microcontrollers. \systName{} provides per-container configurable scheduling and fine-grained access control to host resources through a metadata-driven approach, supporting multiple runtimes via a runtime abstraction layer. We analyze the performance of \systName{} with a small WebAssembly runtime, CS4WAMR, and RIOT OS, a common RTOS. We report on experiments using popular IoT boards based on various Cortex-M microcontrollers. We show the endpoint system brought by \systName{} allowing to regulate access of containers to host resources and provide host services to containers with an overhead of up to 4 ms per call. In particular, we showcase a TinyML use case, whereby containers retain data and model weights, while model inference is delegated to native host RTOS services. 
\end{abstract}



\maketitle

\begin{acks}
The research leading to these results partly received funding from the \grantsponsor{sponsor-MESRI}{MESRI}{dx.doi.org/10.13039/100012948}-\grantsponsor{sponsor-BMBF}{BMBF}{dx.doi.org/10.13039/501100002347} German-French cybersecurity program under grant agreements no \grantnum{sponsor-MESRI}{ANR-20-CYAL-0005} and \grantnum{sponsor-BMBF}{16KIS1395K}. The research leading to these results is also funded by \grantsponsor{501100003032}{ANRT}{http://dx.doi.org/10.13039/501100003032} Convention Cifre n°\grantnum{501100003032}{2024/0426}. Some work was also partially financed by France
2030 via the PEPR Future Networks project FITNESS, and
the PTCC project PQ-OTA. ANRT, MESRI and BMBF are not responsible for any use that may be made of the information this paper contains. This paper reflects only the authors’ views.
\end{acks}

\section{Introduction}

Recent works, such as Aerogel~\cite{liuAerogelLightweightAccess2021} or CS4WAMR~\cite{builTinyMLServiceMultiTenant2025}, have focused on software containerization on microcontrollers by using small WebAssembly (Wasm) runtimes. Containerization enables multiple stakeholders to share a microcontroller -- a scenario called \emph{multi-tenancy}. The challenges of multi-tenancy on microcontrollers are multiple.

The first challenge is fitting the extremely small resource constraints of microcontrollers. To grasp this challenge, readers are referred to RFC7228~\cite{rfc7228} which describes the characteristics of low-end \ac{IoT} devices: a \ac{RAM} budget smaller than 256 KiB, Flash memory smaller than 500 KiB, and CPU clock speeds in MHz.

Another challenge on multi-tenant devices is isolating the different applications running on the device. Isolating memory between applications is common in existing containerization runtimes and is built-in by design in WebAssembly. However, temporal isolation and management of access to host resources are not common and remains difficult~\cite{buil2025cnsm}. 

Very recently, IETF began fostering work in this domain, as demonstrated by the last T2TRG workshop on Composable Code for Things~\cite{t2trg-composable}.
Prior work described in~\cite{wasm-driver-2025} studies I2C and USB peripheral drivers sandboxed in different Wasm components cooperating on Cortex-M devices (Raspberry Pi Pico).
Another example is KaOS~\cite{skrivankova2025kaos}, which introduces a concept to retrofit software using multiple Wasm components on legacy IoT sensor hardware.
Aerogel~\cite{liuAerogelLightweightAccess2021} tackles permissions and energy management for WebAssembly on microcontrollers by proposing to modify WebAssembly Micro Runtime (WAMR) to allow configuring access to peripherals of the device, Aerogel does not allow dynamic loading of permissions with the container and does not allow to have devices with multiple configuration of peripherals.
None of these approaches tackle fine-grained permissions to varied resource types and per-container scheduling on containerization runtimes.
This paper thus proposes a container management framework which fills this gap.

\ \\
\textit{Challenges —} In more detail, we focus on two major concerns regarding runtimes for multi-tenant microcontrollers.

{ \bf 1. Scheduling of containers.} Different applications often require different scheduling priorities. For instance, some applications require more execution time than others. Many runtimes, such as Aerogel~\cite{liuAerogelLightweightAccess2021} or CS4WAMR~\cite{builTinyMLServiceMultiTenant2025}, use the underlying OS scheduler to schedule the different containers. This restricts the capacity to have custom scheduling configurations for each runtime. 

{ \bf 2. Access control for host resources.} When a resource (e.g. a native function) is exposed to one container, it is not possible to limit its exposure from other containers running in the same environment.
Therefore, deploying containers with different permissions requires modifying the firmware and flashing the new version on the device to modify the access rights to the resources. This restricts the possibility to deploy co-located containers with different trust levels.



\ \\
{ \it Contributions — } 
In this paper, we introduce \systName{}, a new container management framework targeting microcontrollers. More specifically:

\begin{enumerate}
    \item we design \systName{}, which provides container lifecycle management, container scheduling, and host resource access control. \systName{} is designed to support different runtimes through a runtime abstraction, support multiple peripheral types with a driver system, and be configurable with lightweight metadata provided with the containers. 
    \item we implement \systName{} on top of RIOT and apply it to hosting small WebAssembly modules using CS4WAMR. \systName{} implementation is available as open-source code\footnote{Available under LGPL-2.1 license at \url{https://github.com/TinyPART/RIOT/tree/tinycontainer/sys/tinycontainer}}.
    \item we perform benchmarks on popular IoT boards (Arduino and Nordic) based on common microcontrollers (Arm Cortex-M4 and Cortex-M33).
    \item our measurements show that \systName{} introduces an acceptable overhead to add crucial security mechanisms.
    \item we apply \systName{} to a tinyML use case which showcases the capabilities of \systName{}. Especially, we highlight the endpoint system of \systName{} which allows containers to access resources outside the container through an interface, and enables hosts to provide services to containers, such as TinyML services.
    \todoBB[later,done]{need to say in a few words which capabilities}
\end{enumerate}

\section{Background} 

\paragraph{Containerization on microcontrollers} 

WebAssembly (Wasm) is a compiled binary format intended for secure and isolated execution within WebAssembly virtual machines. WebAssembly (Wasm) is designed to be compiled from different programming languages. The binary produced after compiling code to WebAssembly is called a module.
WebAssembly can run on microcontrollers through lightweight runtimes such as WebAssembly Micro Runtime (WAMR) \cite{BytecodeallianceWasmmicroruntimeWebAssembly} and Wasm3~\cite{wasm3wasm32019bis}.
On microcontrollers, WebAssembly Micro Runtime (WAMR) supports two execution modes: interpreter mode and AoT (Ahead of Time) compilation mode.
In interpreter mode, WAMR interprets the Wasm bytecode on the device, causing overhead.
AoT compilation involves translating the Wasm bytecode into machine code specific to the target hardware and deploying the AoT code on the IoT device. AoT compilation mode is used to achieve performance close to native execution while keeping isolation guarantees of WebAssembly.
\todoBB[done]{Talk about WebAssembly bytecode interpretation and AoT optimization }

\paragraph{Multi-tenant microcontrollers use cases} 


\citet{builTinyMLServiceMultiTenant2025} highlight commercial solutions like MicroEJ and Atym which offer containerization for microcontrollers. Atym focuses on containers to simplify debugging and development, provide application isolation for security, and support containerized Edge AI. MicroEJ offers a multi-application runtime for sandboxed environments, promoting servitization to allow third-party service development, akin to an app store for microcontrollers. This enhances the user experience and enables device manufacturers to monetize their products with premium features. MicroEJ's applications span smart homes, smart grids, industrial uses, and medical devices.

\paragraph{Fine-grained access control and temporal isolation}
\todoBB[later,done]{à reprendre, mettre dans les paragraphe précédent ce qui ne se différencie pas}
While Wasmtime~\cite{WasmtimeBytecodealliance2017}, one of the main runtimes for WebAssembly, is not adapted to the most constrained devices, other runtimes like WebAssembly Micro Runtime (WAMR)~\cite{BytecodeallianceWasmmicroruntimeWebAssembly} and Wasm3~\cite{wasm3wasm32019bis} are specialized to such targets.
Several propositions modify WAMR to add new functionalities, such as Aerogel~\cite{liuAerogelLightweightAccess2021} which adds energy and permission management, and CS4WAMR~\cite{builTinyMLServiceMultiTenant2025} which adds permission management as well as segmentation and control of memory usage.
Existing propositions do not manage fine-grained access permissions on custom host resources. For example, CS4WAMR uses hard-coded permissions and Aerogel permission system focuses only on peripherals, uses hard-coded drivers, and does not allow loading container permissions along with the container. Moreover, existing Wasm runtimes do not manage the scheduling of container.\\

We propose to use \systName{}, the contribution of this paper, to integrate fine-grained scheduling and permission management on WebAssembly by integrating CS4WAMR as one of the runtimes. We choose CS4WAMR as it is based on WAMR runtime and proposes memory segmentation between containers.



%

\section{\systName{}: secure management of containers on multi-tenant microcontrollers}

\begin{figure}
	\centering
	\includegraphics[width=0.95\linewidth]{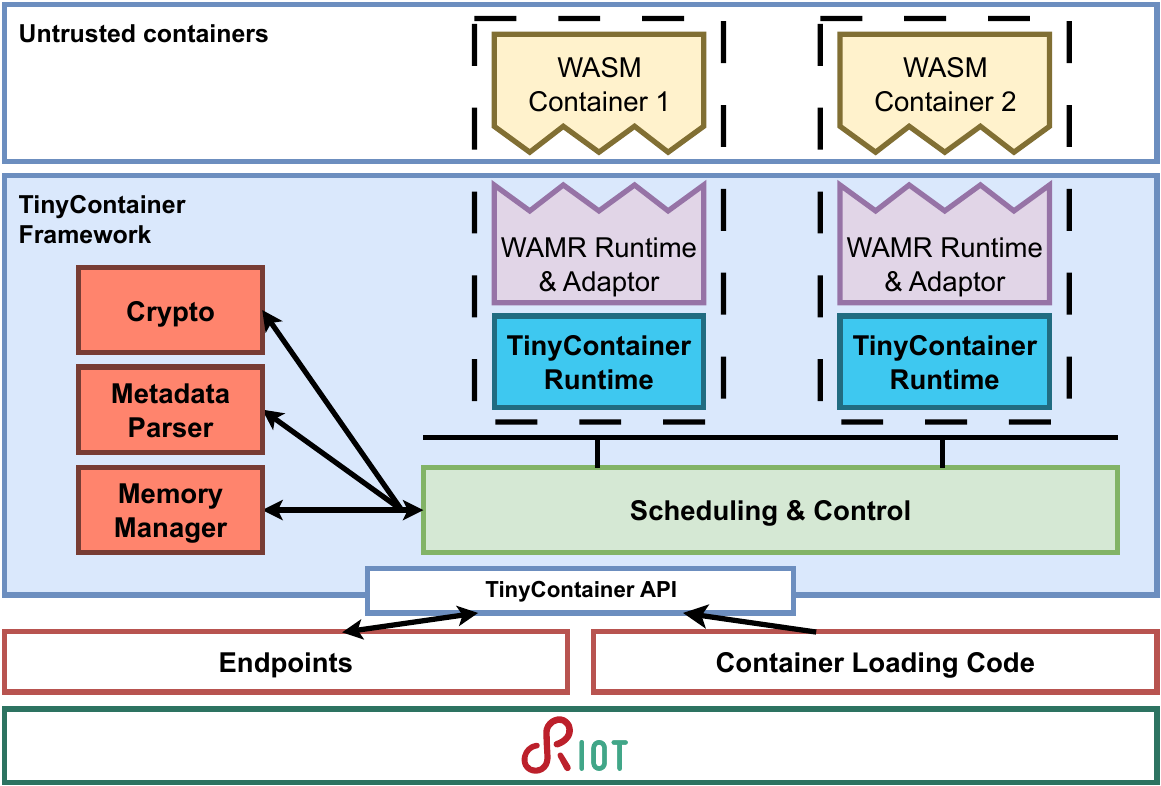}    
    \Description{TinyContainer framework is based on a scheduling and control components. The scheduling and control component runs multiple Wasm containers using a WAMR runtime and WAMR adaptor. TinyContainer exposed multiple API which are used by endpoints and container loading code. }
	\caption{Overview of \systName{} architecture}
	\label{fig:tinycontainer_arch}
    \todoBB[done]{To modify: it can be good to separated \systName{} from the attacker model
    Another improvement path is to put long text in subtitles }
\end{figure}

\systName{} is a container management framework designed for remote management of multi-tenant microcontrollers with built-in security through a per-container configurable scheduler and an endpoint and permission system.
\systName{} is designed to be runtime independent, through an abstraction of the underlying runtime. 
\systName{} uses metadata to control container scheduling and permissions, and checks the integrity and authenticity of containers.

\todoBB[later,done]{differenciating points : endpoints, lifecycle management, isolation for multi-tenancy}


Figure \ref{fig:tinycontainer_arch} provides a high-level view of the architecture. 
The \textit{Scheduling \& Control} component manages containers and exposes its API for container for adding endpoints and managing containers. The runtime abstraction is ensured by the adaptor design pattern through the \textit{Runtime Adaptor} and the \textit{TinyContainer Runtime}. For \systName{} to support a new runtime environment, only the \textit{Runtime Adaptor} component needs to be developed to take into account the specificities of the runtime to be added. The \textit{TinyContainer Runtime} component manages the lifetime of containers. Finally, the \textit{Scheduling \& Control} component uses multiple small components, such as the \textit{Metadata Parser} component to parse metadata file, the \textit{Crypto} component to verify integrity and authenticity of both containers and metadata, and the \textit{memory manager} component to store containers and metadata.

While \systName{} supports multiple runtimes like CS4WAMR \cite{builTinyMLServiceMultiTenant2025}, rBPF runtime from FemtoContainer paper~\cite{zandbergFemtocontainersLightweightVirtualization2022}, and JerryScript, only CS4WAMR runtime adaptor currently supports all the described features. For example, other runtime adaptors currently lack the implementation of some features such as the endpoint system. This is why the paper will mainly focus on the usage of the CS4WAMR WebAssembly runtime.

\todoBB[done]{Talk about remote deployment and loading at runtime}
Containers can be dynamically loaded in \systName{} from a container code file, like a WebAssembly module, a metadata file, and optionally from a data file accessible to the container. To remotely load containers, a typical solution is to use a CoAP server, installed on the device, as a \textit{container loading code} on which clients can send container code and metadata. The \textit{container loading code} can then load containers using the TinyContainer API. The responsibility for ensuring the integrity and authenticity of the code and metadata can then be delegated to \systName{}, which will verify them during container loading.

\todoBB[later,done]{High level code organisation}
\todoBB[later,done]{Figure of the architecture}
\todoBB[later,done]{Put as an introduction to section 3 ???}

\subsection{Access management for containers to diverse resource types}

\begin{figure}[t]
  \centering
  \includegraphics[width=\linewidth]{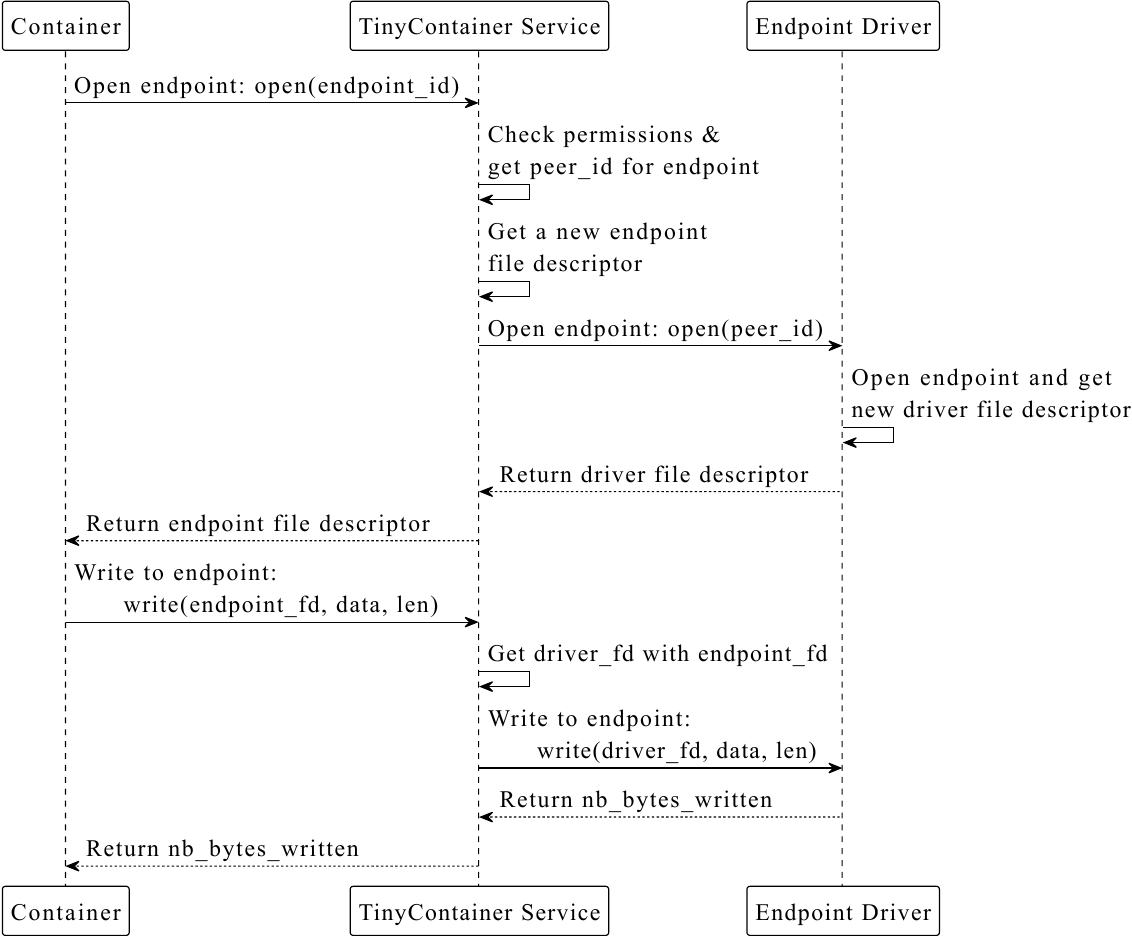}
  \Description{Container opens endpoint to TinyContainer Service with endpoint id. TinyContainer Service checks permissions, get peer if for endpoint and get a new endpoint file descriptor. TinyContainer Service opens endpoint to Endpoint driver with peer id. Endpoint driver opens endpoint and get a new driver file descriptor. Endpoint driver returns driver files descriptor to TinyContainer Service. TinyContainer Service returns endpoint file descriptor to container. The container can then write endpoint to TinyContainer service using endpoint file descriptor, which can get the associated driver file descriptor and write to endpoint driver.}
  \caption{Connection of endpoints from containers up to drivers}
  \label{fig:endpoints_connection}
\end{figure}

\systName{} allows containers to access resources outside the container through an endpoint mechanism. The endpoints can be peripherals, remote services, or services provided by the host to the containers. To control permissions to access endpoints, each container is deployed with metadata declaring container characteristics and the different endpoints accessible by the container.
Container access permissions to endpoints are loaded by parsing the metadata manifest when loading the container.

The endpoint system works similarly to Unix input-output system calls. Containers have access to multiple functions to interact with endpoints. Containers can open a file descriptor to an endpoint using the \textit{open()} function and specifying in argument the target endpoint. When opening the file descriptor, a check is done to verify whether the container has access to the endpoint. If the container does not have access to the endpoint, the opening of the file descriptor fails and the \textit{open()} function returns \textit{-1}. This allows containers to provide fallback mechanisms when not having access to the endpoints due to lacking permissions or lacking peripheral. After obtaining a file descriptor, containers can read and write data to the endpoint using the \textit{read()} and \textit{write()} functions, and finally close the descriptor using the \textit{close()} function.

A typical interaction with a driver is illustrated in \autoref{fig:endpoints_connection}. Multiple identifiers, which are integers, are used by the endpoint system. The \textit{endpoint\_id} is the identifier of the endpoint used by containers to communicate with it. The \textit{peer\_id} is the identifier of the driver used by TinyContainer. The endpoint file descriptor (\textit{endpoint\_fd}) and the driver file descriptor (\textit{driver\_fd}) are respectively the identifier for the communication between containers and TinyContainer Service, and the identifier for the communication between TinyContainer Service and endpoint driver. 
The link between the endpoint and its \textit{endpoint\_id} used by the container, and the link between driver and its \textit{peer\_id} used by driver code are defined in the metadata. This allows container code to be independent of the attribution of the endpoint identifier by the host system. 

The host integrating TinyContainer manages the endpoints, defines the \textit{peer\_id} associated with the different endpoints and provides drivers which implements the \textit{open()}, \textit{read()}, \textit{write()} and \textit{close()} functions to provide access to host services and peripherals to containers.


\todoBB[done]{
Talk here about endpoint id, multiple types of file descriptor, how endpoints and containers are linked together, how to add endpoint.}

\todoBB[done]{

- What is the file descriptor, how are they attributed

- How a driver is implemented.

}

\todoBB[done]{
- WHen checks are done.

- How can container fallback when a permission is missing
}


\todoBB[later,done]{give subportion of CDDL (in annex ?)}

\subsection{Per-container configurable scheduling}

\begin{figure}[t]
  \centering
  \includegraphics[width=\linewidth]{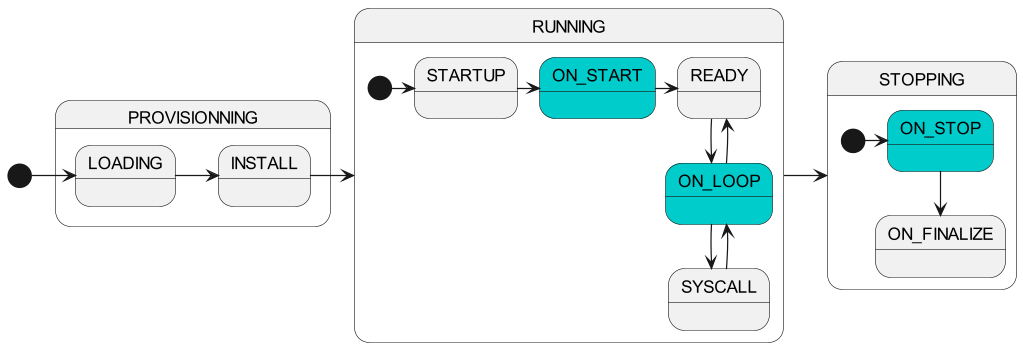}
  \caption{\systName{} container states machine}
  \Description{TinyContainer states machine work in three mains parts: provisionning, running, and stopping. The provisionning contains 2 states: loading and install. The running contains 6 states: startup, on_start, ready, on_loop, syscall. The stopping contains 2 states: on_stop, and on_finalize.}
  \label{fig:container_state_machine}
\end{figure}


A specificity of \systName{} is its ability to abstract the runtime environment, enabling the utilization of identical APIs for initializing, loading, starting, stopping a container, or checking its status irrespective of the runtime in use on a specific IoT device. This feature facilitates a remote provisioning platform in managing numerous devices with varying runtimes remotely. 
As depicted in \autoref{fig:container_state_machine}, the container lifecycle is managed through three distinct entry points corresponding to \verb|ON_START|, \verb|ON_LOOP|, and \verb|ON_STOP| states. A container runtime adaptor must implement the associated entry point functions (\verb|on_start()|, \verb|on_loop()|, and \verb|on_stop()|) to execute container code for the lifecycle.
For instance, in our CS4WAMR implementation, WebAssembly code must expose \verb|start()|, \verb|loop()|, and \verb|stop()| functions, and CS4WAMR runtime adaptor entry points call the respective Wasm functions. 


\systName{} incorporates a control mechanism to control container execution time by implementing a semi-cooperative scheduling system.  
The frequency at which the \verb|on_loop()| function is called is determined by the  container's metadata, using the \textit{loop\_call\_period} value. It is also possible to limit the number of calls to the \verb|on_loop()| function using \textit{loop\_max\_calls} value, after which the container is stopped. 
The maximum execution time is also specified in the container's metadata, with the \textit{start\_exec\_limit}, \textit{loop\_exec\_limit}, and \textit{stop\_exec\_limit} values, ensuring that no single container can monopolize system resources indefinitely. Before executing a container function, a watchdog is armed to be triggered after associated execution time limit. When a container does not respect its execution time limit, the watchdog gets triggered and the execution is returned to \systName{}, which kills the container by stopping its associated thread. This parameter can be adjusted to balance the needs of long-running tasks with the overall responsiveness of the system.

These configurations allow for fine-tuning the container's execution pattern based on its specific requirements and the overall system load. With the timing parameters from the container's metadata, \systName{} provides a flexible framework that can accommodate a wide range of container behaviors while preserving system integrity.

\subsection{Metadata for multi-tenant microcontrollers}

\begin{figure}
	\centering
	\includegraphics[width=\linewidth]{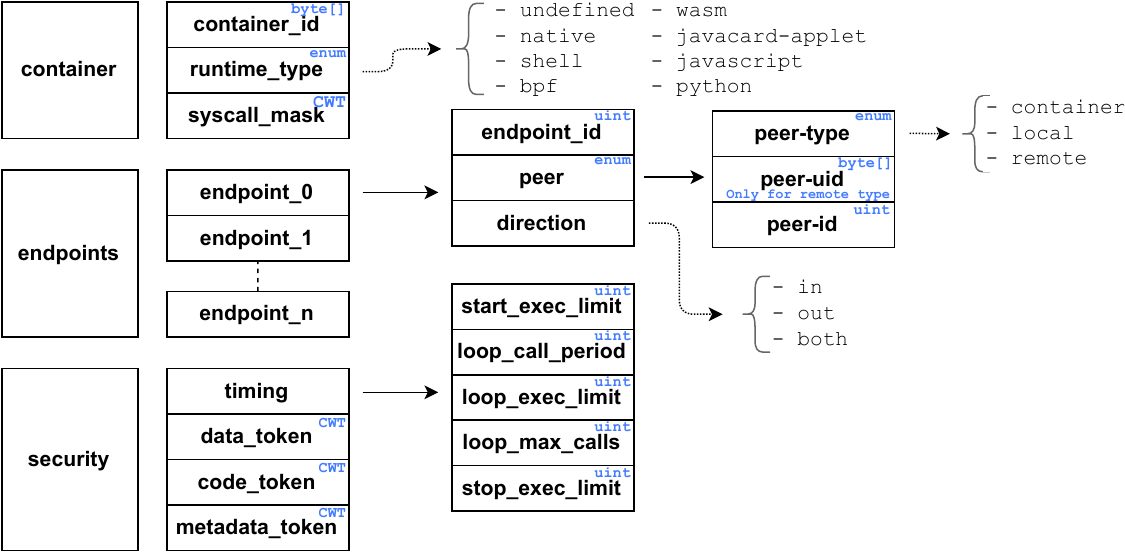}
	\caption{Container Metadata}
	\label{fig:metadata}
    \Description{TinyContainer metadata are divided into three main sections: container, endpoints, and security. The container section includes container ID, runtime type, and syscall mask. The endpoints section lists multiple endpoints, each defined by an endpoint ID, peer information, and a direction. Peer information includes peer type, peer unique ID, and peer ID, with peer types such as container, local, or remote. The security section contains timing, data token, code token, and metadata token. Timing contains execution limits and periods, which are start execution limit, loop call period, loop execution limit, maximum loop calls, and stop execution limit. }
    \todoBB[done]{Update image with new format, remove authorization code}
\end{figure}


To build metadata ready for transfer over low-power networks and with built-in security in the format, we use CBOR (Concise Binary Object Representation) and COSE (CBOR Object Signing and Encryption) as used by the SUIT manifest~\cite{moranConciseBinaryObject2025}, a state-of-the-art manifest format for remote firmware deployment on an Internet of Things devices. CBOR is a data format similar to JSON but designed to have a minimum size for constrained devices and network exchange. COSE is a specification to create signatures of CBOR data. CBOR Web Token (CWT) \cite{jonesCBORWebToken2018} allows creating CBOR-based tokens containing both a COSE signature and the signed data. \systName{} metadata uses the CBOR format to encode metadata with minimum overhead and CWT to sign permissions of the container (syscall\_mask), the hash of WebAssembly container, the hash of container data, and the hash of the metadata itself. 

As depicted in \autoref{fig:metadata}, the metadata consists of three sections: containers description, endpoints description, and security-related description. The container description contains a container identifier, the type of container (e.g. Wasm, rBPF) and a signed token which contains a byte map describing the list of endpoint calls authorized for this container. The endpoint description contains a list of endpoints characterized by an ID, a type, and an access direction, that lists which endpoint functions are available to be used by the current container. As illustrated with the fields "...", the structure is flexible enough to add an unfixed number of fields to describe the endpoints.
The security-related description contains a detailed description of limits related to the container lifecycle, as well as three tokens to verify the authenticity and integrity of data, code, and metadata.

The Container Metadata contains four CBOR Web Tokens (CWT), each generated by different roles which may be involved in the container's lifecycle. The software developer creates the code token, while a service provider can add data to customize the code supplied by the developer and create the associated data token. The deployment infrastructure manager is responsible for generating the metadata token. Lastly, the owner of the microcontrollers produces the syscall mask token to authorize the use of specific resources on its device by the container. Managing this collection of tokens falls under the responsibility of the container management platform and is beyond the scope of this paper. It should be noted that all these roles can be carried out by a single actor or by multiple actors.

\todoCG[done]{Explain what security mechanism has been implemented (cut and past from metadata + info ?) Multiple COSE signature as different part of the metadata could be provided by multiple different actors. Like an actor providing code, another producing metadata and deploying container (data can be provided by eithers)}

\todoBB[later,done]{Talk about CBOR WebToken}
\todoBB[later,done]{Why is it logic in the article to have metadata cwt with endpoints and  }


\section{Security considerations} 

\paragraph{Security definition}

\begin{figure}
    \centering
    \includegraphics[width=0.9\linewidth]{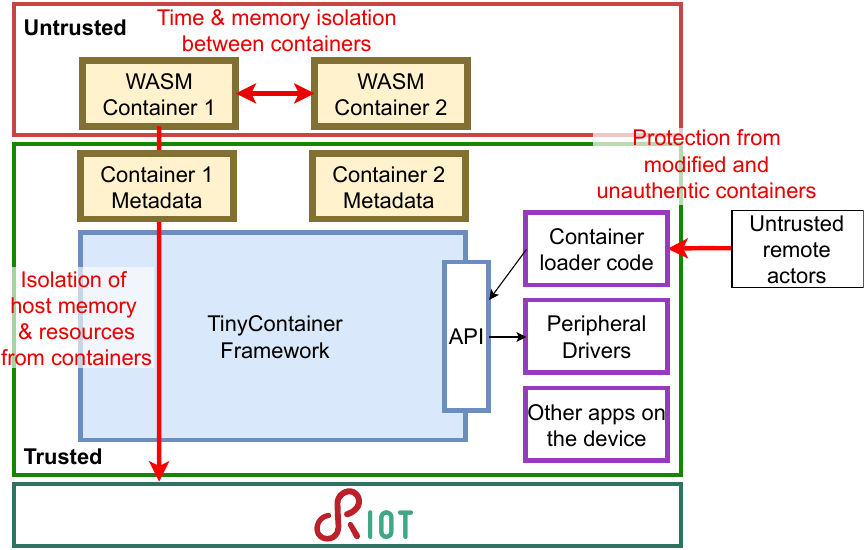}
    \Description{TinyContainer security model supports time and memory isolation
between containers, isolation from containers of host memory and resources, and protection from modified and unauthentic containers deployed by untrusted remote actors. }
    \caption{\systName{} security model}
    \label{fig:security-model}
\end{figure}

\systName{} provides sandbox isolation, as defined by \cite{lefeuvreSoKSoftwareCompartmentalization2024}, which consists of reducing container permissions to protect the host system, and the other containers. Contrary to technologies like ARM Trustzone, Sancus, \systName{} does not provide safebox isolation, which consists of reducing the privilege of the system to access a container. Thus, \systName{} allows the execution of mutually distrusting containers, which entrust the host and \systName{} to provide sufficient resources for their execution. Containers having different trust levels means that containers are not trusted equally to access resources of the system and therefore have different permissions to access the systems. \systName{} security model is represented in \autoref{fig:security-model}.

\paragraph{Security hypotheses}

To maintain these security guarantees, multiple security hypotheses must be respected.

The runtime used must provide both fault and memory isolation. WebAssembly was explicitly designed with these isolation guarantees in mind. However, WebAssembly bytecode interpreters require the bytecode to be validated. While some runtimes, such as WAMR~\cite{BytecodeallianceWasmmicroruntimeWebAssembly}, perform the validation on-device, others, like Wasm3~\cite{wasm3wasm32019bis}, assume that validation has already been carried out, potentially off-device. 
Some runtimes, like WAMR\cite{BytecodeallianceWasmmicroruntimeWebAssembly} and Wasmtime~\cite{WasmtimeBytecodealliance2017}, support ahead-of-time compilation. To preserve equivalent security properties, AoT requires that the Wasm-to-native-code AoT compilation be done correctly by a trusted party. \systName{} security holds as long as the Wasm security guarantees are upheld. Thus, \systName{} is compatible with both interpreted and AoT, but for interpreters relying on off-device validation and for runtimes using AoT, \systName{} requires that the trusted entity signing the metadata both validates the Wasm bytecode, and in AoT case, performs the AoT compilation. 
Another security requirement is the correct implementation of the endpoint to maintain isolation. For instance, in the following section, we implemented a TinyML driver which allows performing inference from data and model weights while preserving isolation between containers.

\paragraph{Attacker model}

Our attacker model assumes that an attacker has full control over the container code as long as the container passes the validation step. Attackers can load untrusted containers, which can, for instance, try to access peripherals of the device or try to monopolize execution time. An attacker can communicate with the device over the network, for instance to load containers on the device.

Attackers do not have control over the metadata of their container, as metadata are defined by a trusted entity that manages the device. Attackers breaking the cryptographic primitives used in metadata, notably to deploy invalid containers, is out of scope of this paper. Attackers do not have physical access to the hardware.

Other applications trying to access information of the containers is not protected by \systName{}, and thus, is out of scope of this paper.

\section{Application on TinyMLaaS use cases} \label{section:tinyml-usecase}

\systName{} notably enables the host to provide services to containers and manage access control to them with metadata.
We propose to test this mechanism of host-provided services with a use case of Tiny Machine Learning as a Service (TinyMLaaS).

\subsection{Tiny Machine Learning as a Service (TinyMLaaS)}


Doyu et al. \cite{doyutinymlaasecosystemmachine2021} introduce TinyMLaaS, a solution involving a Cloud or Edge platform that uses model compilers for efficient ML model compilation and deployment on microcontrollers. WASI-NN~\cite{WebAssemblyWasinnNeural2025} is a proposal for WebAssembly System Interface (WASI) allowing WebAssembly modules to send TinyML models to the host and run them in machine learning runtimes such as TensorFlow Lite~\cite{davidTensorFlowLiteMicro2021} and OpenVINO~\cite{OpenvinotoolkitOpenvino2025}.
Buil et al.~\cite{builTinyMLServiceMultiTenant2025} propose an architecture of TinyMLaaS with a WebAssembly container by creating AoT-compiled WebAssembly containers from machine learning model files for running model inference.

\todoBB[done]{Talk here about wasi-NN}

\subsection{TinyMLaaS endpoint use case}

CS4WAMR article \cite{builTinyMLServiceMultiTenant2025} highlights the high overhead of Wasm bytecode interpretation for compute-intensive tasks such as TinyML, which makes interpreters impracticable for such tasks or requires the use of Ahead-of-Time compilation which does not have such overhead. 

To address this issue, we propose an architecture where the model weights are in the container, but machine learning inference is delegated to the host. The mechanism uses a predefined model structure, so that model code is already being present in the host, but containers can customize inference by changing the weights.
Our proposition works by having an endpoint dedicated to machine learning which can first receive model weights from the container, then receive the input data for prediction, compute the result, and finally let the container read the result.

This architecture allows to leverage the computation optimizations of the devices, not available when doing bytecode interpretation, but without needing AoT optimization which requires a trusted third party to compile the WebAssembly module to machine code and without requiring having a full model runtime, like LiteRT (former TensorFlow Lite), which uses a lot of memory and storage.
This architecture allows to leverage models using the same structure, such as models that can be fine-tuned and brought by the container, and models using a standardized structure.

A typical usage of the TinyML endpoint involves the device host either defining a structure for accepted models, e.g. a multi-layer perceptron with a defined size for each layer, or furnishing a model that can be customized and fine-tuned by container providers. 


\subsection{Implementation of TinyMLaaS on \systName{}}

TinyML service has been implemented as an endpoint for \systName{}. The TinyML endpoint includes the ML service code to load and communicate with the thread responsible for machine learning processing, as well as the machine learning code, specific to the model structure, used to compute predictions.

\paragraph{Communication between the TinyML endpoint and the container}

\begin{figure}
    \centering
    \includegraphics[width=0.85\linewidth]{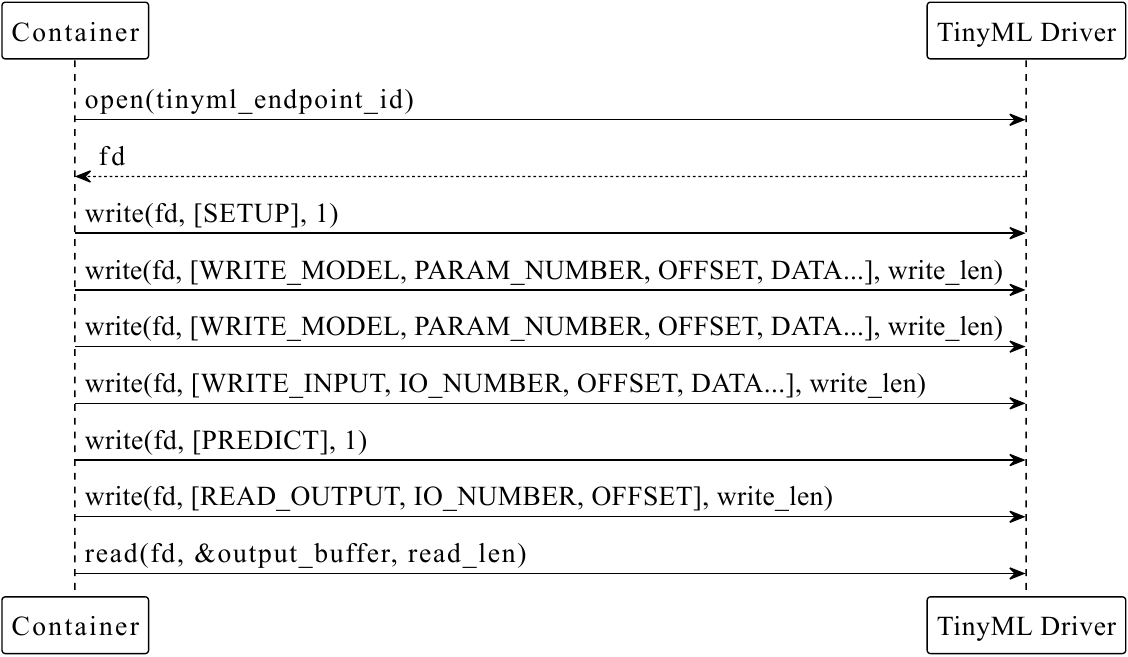}
    \caption{TinyML endpoint usage}
    \Description{This figure shows the sequence of interactions between a container and a TinyML driver. Each step is a call to either open, read, or write function from the container. The process starts with the container opening a TinyML endpoint and receiving a file descriptor using open function. Then using write function, the container then sends a setup command, followed by write model commands. Next, it writes input data, sends a predict command, and requests to read the output. Finally, the container reads the output buffer from the TinyML driver using read function.}
    \label{fig:tinyml-endpoint-sequence}
\end{figure}

Containers can interact with the TinyML endpoint through the endpoint system offered by \systName{}. The sequence diagram in \autoref{fig:tinyml-endpoint-sequence} illustrates the communication steps.
To load the model, two steps are necessary after establishing the connection between the container and the TinyML endpoint. Initially, the container must transmit the model weights to the TinyML endpoint. This enables the TinyML endpoint to reconstruct the model based on the parameters and model structure. Once the model is loaded into the TinyML endpoint, the container can perform predictions by sending a vector and receiving the predicted label. 
\todoBB[later,done]{Todo sequence diagram} 

\paragraph{Generation of TinyML endpoint code}

TinyML endpoint contains the ML code generated from a TensorFlow Lite model, and the ML service code to interact with the model.
The machine learning code is generated from RIOT-ML~\cite{huangriotmltoolkitair2024}, a machine learning toolkit which integrates a model compiler, based on Apache-TVM~\cite{chenTVMAutomatedEndtoend2018}, to generate machine code. Using a script, the code produced is then automatically integrated with the ML service code in the TinyML endpoint.

\section{Evaluation}

\subsection{Methodology}

The evaluation was conducted on the 2025.07 release of RIOT-OS.
All WebAssembly modules were built using clang version 18.1.3 with optimizations, avoiding the static linking of libraries like libc or those related to WASI. To tailor these modules for the limited memory of microcontrollers, linker flags were utilized to minimize the size of Wasm modules by setting the linear memory stack size to 1024 bytes and the global base, which determines the starting address of data in linear memory, to 16 bytes. Additionally, the compilation and linking processes employed the optimization flags \textit{O3} and link-time optimization. The RIOT-ML 82b1b5c commit version and Apache TVM v0.18.0 were utilized. All timing measurements presented in this document were performed using the device's hardware timer.

The benchmarks were conducted on two different boards. The first board is the Arduino Nano 33 BLE (rev1), which features an nRF52840 CPU utilizing an ARM Cortex-M4, running at 64MHz, with 256 kilobytes of SRAM and 1 megabyte of flash memory. The second board is the nRF9160-DK, equipped with an nRF9160 CPU based on an ARM Cortex-M33, also clocked at 64 MHz, and includes 256 kilobytes of SRAM and 1 megabyte of flash.

\subsection{Performance evaluation}

In this section, we evaluate the performance and memory impact of the \systName{} framework. We compare the code running WebAssembly Micro Runtime (WAMR)~\cite{BytecodeallianceWasmmicroruntimeWebAssembly}, CS4WAMR~\cite{builTinyMLServiceMultiTenant2025} and our \systName{} proposition with CS4WAMR runtime.

To conduct our evaluation, we have created WebAssembly containers to evaluate the impact on performances of \systName{}. First, we have created a polling container that reads data from an endpoint at each loop call to measure endpoint access performance. Then, we have created friendly containers executing \textit{NOP} instructions for a small period of time and a malicious container monopolizing execution. Finally, we define a running minimal runnable container which is the smallest possible container which can run on the runtime to evaluate memory usage.

\subsubsection{Metadata and access control overhead}

\autoref{tab:execution-time} summarizes the performance overhead to load a polling container and execute it with WAMR, CS4WAMR and \systName{}. The polling container uses a data endpoint to read two buffers of 1960 bytes each. The polling container works by reading 255 bytes at a time from the endpoint, hence, polling the endpoint to read the two buffers.

Loading a container with \systName{} is highly impacted by the loading of the metadata. This is due to the introduction of several additional steps, namely parsing CBOR objects, retrieving the public key, verifying the validity of the CWTs present in the manifest using the EdDSA algorithm, extracting the hashes of the CWT claimsets, and verifying that the hashes correspond to the metadata, data, and code. In particular, on the nRF9160-DK, the implementation of RIOT-OS does not activate the hardware cryptographic accelerator, so all the calculations were done in software which greatly impacted the performance.


The container execution is also impacted because each driver call must pass through \systName{} endpoint system, which routes the communication to the correct driver. On the Arduino Nano 33 BLE, each \textit{read()} endpoint call to read 255 bytes takes 4 ms with TinyContainer, while taking 11 us with CS4WAMR. 255 bytes per call is currently a limit of \systName{} regarding the number of readable bytes at a time.



\begin{table}
    \small
    \tabcolsep3pt    
    \newcommand{\tabrule}[0]{\rule{12pt}{0pt} }
	\caption{Execution time of WebAssembly runtimes and framework for running a polling container}\label{tab:execution-time}
	\begin{tabular}{@{}llcc@{}}
		\toprule
		     &   & \multicolumn{2}{c}{Execution time (in ms)} \\
        \cmidrule{3-4}
		\multicolumn{2}{l}{Runtimes} & Arduino Nano 33 BLE & nRF9160-DK \\
		\midrule
		\multicolumn{2}{l}{\textbf{WAMR Interpreter}} & \textbf{}  & \textbf{} \\
        \tabrule & Load container & 1.61 & 1.67  \\
         & Container execution & 0.91 & 0.98 \\
		\multicolumn{2}{l}{\textbf{CS4WAMR}}  & \textbf{}  & \textbf{} \\
         & Load container & 2.21 & 2.16 \\
         & Container execution & 0.97 & 0.99 \\
		\multicolumn{2}{l}{\makecell{\textbf{\systName{} with CS4WAMR}\\\rule{8pt}{0pt}\textbf{(without cryptography)}}}  & \textbf{}  & \textbf{} \\    
        & Load container metadata & 19.50 & 19.33 \\
         & Load container & 3.54 & 6.04 \\
         & Container execution & 63.23 & 60.10 \\
		\multicolumn{2}{l}{\textbf{\systName{} with CS4WAMR}}  & \textbf{}  & \textbf{} \\    
        & Load container metadata & 91.15 & 8 810.16 \\
         & Load container & 3.61 & 6.05 \\
         & Container execution & 74.0 & 63.54 \\
		\bottomrule
	\end{tabular}
\end{table}


\subsubsection{Scheduling overhead \& guarantees}


%

\todoBB[later,done]{Measure setup Watchdog cost}

In this section, we evaluate the performance overhead of the scheduling mechanism.

First, we have measured the performance overhead of the watchdog mechanism by measuring the time added by the mechanism. By running an example code on Arduino Nano 33 BLE, we observe that the setup of the watchdogs takes only 3 microseconds before the call of a container function and the killing of a container not respecting its contract takes 4 microseconds.

\begin{figure}
    \centering
    \includegraphics[width=0.99\linewidth]{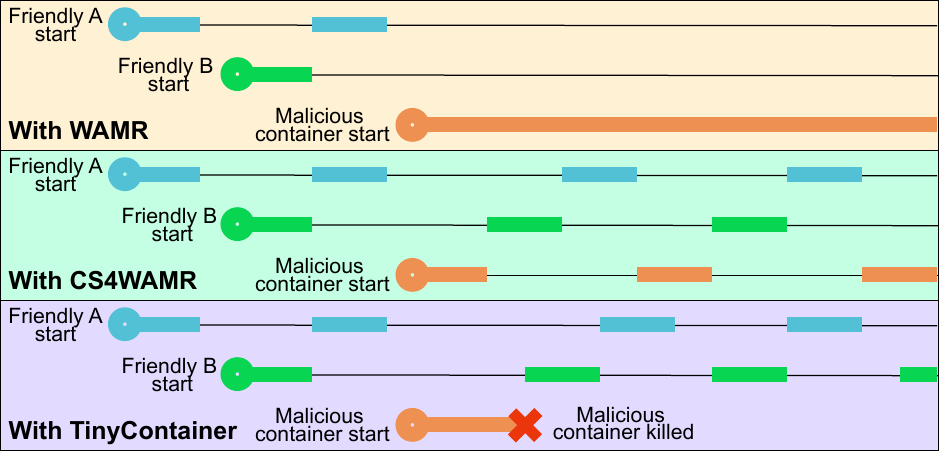}
    \caption{Representation of the scheduling of containers across WAMR, CS4WAMR, and TinyContainer with long-running malicious task}
    \Description{This figure compares the behavior of WAMR, CS4WAMR, and TinyContainer when running two friendly containers and one malicious container monopolizing execution. In the WAMR system, the malicious container runs uninterrupted and blocks execution of the friendly containers. In the CS4WAMR system, the malicious container runs alongside the friendly containers. In the TinyContainer system, the malicious container is started but is killed after some time, allowing the friendly containers to continue to run normally.}
    \label{fig:scheduling-comparizion}
\end{figure}

\todoBB[later, done]{Figure as in demo paper with the timing on the different implementation.}
We have also compared container scheduling with WAMR, CS4WAMR and \systName{} with a container not respecting its time contract. 
In order to do so, on each implementation, we have executed 3 containers: two friendly containers, and one malicious/dysfunctional container. The different scheduling of the containers are represented in \autoref{fig:scheduling-comparizion}. The friendly containers execute for a few milliseconds and return the execution. The malicious or dysfunctional container executes but never returns the execution. As WAMR does not manage scheduling, we run one container after the other and see the malicious container monopolizing execution and blocking other containers from executing. With CS4WAMR, we observe that the malicious container is not killed but preempted, leaving a constant cost of running the malicious container each time it is called then preempted after scheduler timeout. With \systName{}, the malicious container exceeds the execution time limit on its first call and is killed, leaving only the two friendly containers running.


\subsubsection{Memory usage evaluation}

\todoCG[done]{Table 2}

The memory usage of WAMR, CS4WAMR and \systName{} with CS4WAMR to run a minimal container is presented in \autoref{tab:ROM-RAM-tinyml}. The major impact of \systName{} either on RAM or ROM corresponds to the addition of the \systName{} library for a single container. The overhead of \systName{} for each additional container is mainly caused by the memory usage of CS4WAMR which requires a buffer given by TinyContainer. The overhead is also notably caused by the extra 485 bytes of metadata and the additional array elements to store containers and metadata.
\todoCG[later,done]{Discuter de la différence RAM/ROM pour le conteneur additionel parce qu'il y a une différence}

\begin{table}
    \tabcolsep3pt
	\caption{ROM and RAM usage in bytes of WebAssembly runtimes and framework for running minimal runnable container on Arduino Nano 33 BLE}\label{tab:ROM-RAM-tinyml}
	\begin{tabular}{@{}llcccc@{}}
		\toprule
        &       & \multicolumn{2}{l}{For 1 container} & \multicolumn{2}{l}{\makecell{Per additional\\container}} \\
        \cmidrule(rl){3-4} \cmidrule(rl){5-6}
		\multicolumn{2}{l}{Runtimes} & ROM & RAM & ROM & RAM \\
		\midrule
		\multicolumn{2}{l}{\small \textbf{WAMR Interpreter}} & \textbf{}  & \textbf{} \\
        \rule{12pt}{0pt} & Wasm module & 124 & 124 & +124 & +124  \\
        & WAMR library & 43817 & 7056 & +0 & +6969 \\
        & Bootstrap code & 706 & 304 & +60 & +8 \\
		\multicolumn{2}{l}{\textbf{CS4WAMR}}  & \textbf{}  & \textbf{} \\    
        & Wasm module & 124 & 0 & +124 & 0 \\
        & WAMR library & 43821 & 67 & +0 & +0\\
        & CS4WAMR library & 1710 & 1032 & +0 & +0 \\
        & Bootstrap code & 454 & 10104 & +40 & +10008 \\
		\multicolumn{2}{l}{\textbf{\systName{} with CS4WAMR}}  & \textbf{}  & \textbf{} \\    
        & Wasm module & 124 & 124 & +124 & +124 \\
        & WAMR library & 43821 & 67 & +0 & +0 \\
        & CS4WAMR library & 1030 & 128 & +0 & +0 \\
        & \systName{} library & 8798 & 16870 & +156 & +9900 \\
        & Bootstrap code & 2415 & 739 & +862 & +534\\
		\bottomrule
	\end{tabular}
\end{table}





\subsection{TinyML approach comparisons}

\todoCG[done]{Table 3: interpreter slow due to bytecode interpretation, our solution (TinyFontainer): security - communication driver-container}

In the \autoref{tab:inference-time}, we compare the inference time of our TinyMLaaS proposition using TinyContainer endpoint, as described in \autoref{section:tinyml-usecase}, with the TinyML in the container proposed by Buil et al.~\cite{builTinyMLServiceMultiTenant2025}. Performing TinyML inference in-container with a WAMR interpreter has a high cost on the inference time, while the overhead introduced by our solution to ensure the communication between the container and the driver is limited. Delegating computation to the host drastically reduces TinyML inference overhead without doing AoT compilation. 

Thus, delegating to the host compute intensive tasks is a solution to remediate the high overhead of WebAssembly interpreter. TinyContainer enables a way to expose such services from the host to the container.


\begin{table}
    \small
	\caption{Model inference time comparison on Arduino Nano 33 BLE and nRF9160-DK boards}\label{tab:inference-time}
    \renewcommand{\defaultaddspace}{3pt}
	\begin{tabular}{@{}cccc@{}}
		\toprule
		       & \multicolumn{2}{c}{\makecell{Inference time (in ms) with\\DS-CNN Small model~\cite{zhangHelloEdgeKeyword2018}}} \\
        \cmidrule{2-3}
                & Arduino Nano 33 BLE & nRF9160-DK  \\
		\midrule
		\makecell{In-container\\WAMR AoT}          & 442 & 497 \\
        \addlinespace
		\makecell{In-container\\WAMR Interpreter}  &  48 324   & 48 130 \\ 
        \addlinespace
		\makecell{\systName{} TinyML\\native endpoint}  &   512  & 507 \\ 
		\bottomrule
	\end{tabular}    
    \\
    {    
    \scriptsize \raggedright The inference time is the measured time of the execution of the container function running inference or the container function calling inference endpoint for TinyContainer. It includes all the inference time and the communication cost, and the Wasm container function starting cost. 
    For TinyContainer part, it does not include the loading of the model from the container to the driver.
     \par }
\end{table}


\section{Discussions}

\subsection{Limitation of endpoint model}

In the endpoint model used by \systName{}, containers should be proactive in data exchange. A container cannot be called by an endpoint on the device. Therefore, to obtain updates, a container should pull updates from the endpoints. This might limit the possibility of the device being placed in sleep mode to save energy. One feature for future work is to allow containers to be put in sleep mode and be woken by endpoints. Then the container could pull from the endpoint to get information about the update.

Another limitation of current implementation is the need for hosts to implement endpoints that could be used by containers, as the current implementation does not provide built-in endpoints to access device peripherals. One solution to fix this issue can be by adapting existing works that provide interfaces to containers like WASI systems. 

\todoBB[later,done]{(but there are limitation, limit of different permission per containers...)}

\subsection{Advantages and limitations of \systName{} scheduling} 

The scheduling model proposed with \systName{} is cooperative scheduling with a watchdog timer. This allows to have no unwanted context switching producing overhead as containers can choose when to return control to \systName{}. \systName{} scheduler automatically kills containers not respecting their contract defined in the container metadata. This prevents malicious or buggy containers from having a significant impact on other containers.

The current scheduling model, while customizable per container, is constant over time. Consequently, it does not allow mechanisms like Cloud bursting, which would enable containers to use unused resources when the device is under low utilization.

Another challenge is the definition of loop execution time limit (\textit{loop\_exec\_limit}) for each container, as a value too small puts the container at risk of being killed by the watchdog. Therefore, users might overprovision this value, and thus increase the maximum possible period of time that a container could wait to be executed. 

\subsection{Multi-tenant metadata overhead} 

\todoBB[done]{Talk about overhead of having multiple signature and provide alternative: like a single verification and not multiples}
\todoBB[later,done]{Talk about data file to talk about card personalization}

\systName{} metadata are conceived to be constructed with information from multiple actors. For example, one actor creates the container code and signs the code. Another actor personalizes the container by setting container data and signs the data. Finally, the entity managing \systName{} aggregates all the signatures, produces container metadata, and signs the metadata. 

This system allows reducing possible interference between actors, allowing better understanding of the root cause of a problem in case of issue. For example, if a bug causes an issue in a container, \systName{} gives the guarantee that the code of the container has not been modified by any other actor, as the code can be signed by the entity producing the code, and thus, the actor producing container code is potentially responsible. Another advantage is that if an actor uses a third-party provider using \systName{} to remotely deploy its container on its device, the actor is assured that only containers that it has signed can be deployed on the device.

One major drawback of this system is the need to verify multiple signatures when loading the container, which produces high overhead on boards not having a cryptographic processor. One alternative, if the metadata does not need to have the properties described above is to only keep the signature for the \textit{metadata\_token} or even to use only the \textit{metadata\_token} with Message Authentication Code (MAC) but with the requirement of using one unique key per device.

\subsection{Advantages and limitations of host-delegated TinyML}

Delegating TinyML to the host using \systName{} allows for lighter machine learning implementations, containing only the required code, which is more suited for constrained microcontrollers compared to solutions like wasi-nn, which requires a model runtime like TensorFlow Lite which often means a non-negligible overhead. Additionally, deployed containers do not need Ahead of Time compilation to achieve good performance, compared to solutions such as \cite{builTinyMLServiceMultiTenant2025}, which require a trusted server for AoT compilation. However, this approach requires preinstalling machine learning code on the device, so a device can only support a single model structure at a time.

Another constraint with this solution is the need to trust the host to give him the model weights, which can be a business value, to do the machine learning computation. This limitation can be acceptable as the containerization solutions used provide a sandbox, where the containers are protected from other containers but not from the host, and not a safebox, where the containers are protected from the host.



\section{Related Works} 

\subsection{Runtimes}

\begin{table*}
    \newcommand{\yescase}[0]{
        \makecell{\normalsize\textcolor{ForestGreen}{\ding{51}}}
    }
    \newcommand{\nocase}[0]{
       \makecell{\normalsize\textcolor{Bittersweet}{\ding{55}}} 
    }
    \centering
    \tabcolsep2pt
    \renewcommand{\arraystretch}{1.5}
    \caption{Feature comparison between containerization and runtime solutions}
    \label{tab:runtimes-comparition}
    \footnotesize
    \begin{tabular}{cccccccc} 
        \toprule
        Runtimes & Bytecode &  \makecell{Permission\\management\\per container} & \makecell{Access-control\\metadata\\format} & \makecell{Access to\\multiple types of\\host resources} & \makecell{Configurable\\execution time\\per container} & \makecell{Support\\AoT-compiled\\containers}  \\ 
        \midrule
        \makecell{Femto-\\Containers~\cite{zandbergFemtocontainersLightweightVirtualization2022}} & \makecell{ebpf-\\based} & \yescase & \nocase & \yescase & \nocase & \nocase \\ 
        VeloxVM~\cite{tsiftesVeloxVMSafe2018} & Custom  & \yescase & \yescase & \yescase & \yescase & \nocase \\ 
        \makecell{Polyglot\\Cerberos~\cite{akkermansPolyglotCerberOSResource2018}} & Custom & \yescase & \yescase & \yescase & \yescase & \nocase \\ 
        \makecell{Toit~\cite{ToitHighlevelSoftware}} & Custom & \nocase & \nocase & \yescase & \yescase & \nocase \\
        Aerogel~\cite{liuAerogelLightweightAccess2021} & Wasm & \yescase & \yescase & \nocase & \nocase & \nocase \\ 
        WAMR~\cite{BytecodeallianceWasmmicroruntimeWebAssembly} & Wasm & \nocase & \nocase & \yescase  & \nocase & \yescase \\ 
        Wasmico~\cite{ribeiroWASMICOMicrocontainersMicrocontrollers2024} & Wasm & \yescase & \nocase & \yescase & \nocase & \nocase \\ 
        \makecell{CS4WAMR} \cite{builTinyMLServiceMultiTenant2025} & Wasm & \yescase & \nocase & \yescase & \nocase & \yescase \\ 
        
        \makecell{\systName{}} & Multiples & \yescase & \yescase & \yescase & \yescase & \yescase\\ 
        \bottomrule
    \end{tabular}
\end{table*}

WebAssembly (Wasm) is easily associated with containers as it is based on Wasm modules, which are runnable units of software, and runtimes like Wasmico \cite{ribeiroWASMICOMicrocontainersMicrocontrollers2024} propose to manage WebAssembly modules as containers with features like containers pause and resume, test \& development, delivery, and remote operation by exposing an HTTP server for remote management and monitoring.

Several efforts focus on integrating additional control in WebAssembly runtimes on microcontrollers, either through permissions control~\cite{liuAerogelLightweightAccess2021, builTinyMLServiceMultiTenant2025}, energy control~\cite{liuAerogelLightweightAccess2021}, and memory consumption control~\cite{liuAerogelLightweightAccess2021,ribeiroWASMICOMicrocontainersMicrocontrollers2024,builTinyMLServiceMultiTenant2025}.

The WebAssembly component model \cite{WebAssemblyComponentModel} is a proposal for WebAssembly specification to implement a format that describes container permissions and container composition, making inter-container communication interface easier. But both the Component Model, which is only available on a component-aware runtime such as Wasmtime or on a component layer, such as wasm\_component\_layer \cite{dwyerDouglasDwyerWasm_component_layer2025}, induce a costly overhead for constrained microcontrollers.
\citet{liuAerogelLightweightAccess2021} propose Aerogel to have an access-control specification provided with the container, as in the Wasm component model, but adapted to the most constrained devices. 

While runtimes can manage permissions to host resources, hosts should provide access to system resources to containers. To do that, WASI (WebAssembly System Interface) \cite{dangohmanwebassemblywasiv0242025} defines a standard interface to access system resources. \citet{wasm-driver-2025} propose a WASI proposal to allow containers to interact with I2C and USB, often used on IoT devices. Aerogel~\cite{liuAerogelLightweightAccess2021} proposes its own system to access memory-mapped peripherals.



Runtimes using other bytecode or techniques exist on microcontrollers. For example, Toit~\cite{ToitHighlevelSoftware} provides containers with a memory-safe language providing fault isolation for robustness and devops-style management of containers with over-the-air deployment.
But, Toit is not made for mutually distrusting containers, as a malicious container can impact the execution of other containers, as any container can use the full Toit API, including the system API to control containers and networks. Femto-containers~\cite{zandbergFemtocontainersLightweightVirtualization2022} uses rBPF-based containers that provide isolation and updatability with very small footprints on RIOT OS, 
VeloxVM~\cite{tsiftesVeloxVMSafe2018} and Polyglot Cerberos~\cite{akkermansPolyglotCerberOSResource2018} use custom bytecodes with custom schedulers to orchestrate containers on microcontrollers with timing control.

At the operating-system level, Tock OS~\cite{levyMultiprogramming64kBComputer2017} combines language‑based isolation for kernel “capsules” and hardware‑enforced user processes using Memory Protection Unit (MPU), giving fine‑grained isolation with predictable footprint. It uses a preemptive scheduler for processes (round‑robin by default) and cooperative scheduling inside the single‑threaded kernel event loop.

\todoBB[done]{Add about tock os and say that it does not allow ??}

We compare \systName{} implementation with CS4WAMR with runtimes from state of the art in \autoref{tab:runtimes-comparition}.

\subsection{TinyML from containers on microcontrollers}

\todoCG[done]{section 7.2 EWSN; to talk about TinyML on microcontrollers globally. is more SOTA on tinyml, if yes take from CS4WAMRpaper (tinymlops, and tinymlaas)}
\citet{builTinyMLServiceMultiTenant2025} summarize existing approaches in MLOps and TinyMLOps while highlighting the benefits of using WebAssembly to enhance flexibility and modularity.

Prior to our proposition, two main methods existed to run machine learning models from WebAssembly containers on microcontrollers.

The first one, WASI-NN~\cite{WebAssemblyWasinnNeural2025}, works by having the container provide the model file to the host that run inference using a model runtime, such as TensorFlow Lite Micro, OpenVINO, and ONNX Runtime. This method allows the device to run diverse model types, but requires a full model runtime on the device.

The second one proposed by \citet{builTinyMLServiceMultiTenant2025} compiles the model file into machine code and puts it in a WebAssembly container. This allows to not require any code from the host, but requires AoT compilation to have good performance, as bytecode interpretation has a significant overhead.

Our proposition of having preinstalled machine learning code on the host is a mix of the two methods, having both the efficiency of the first method even with bytecode interpretation and the lightness of the second method. The limitations are that the model types on one device are fixed and offer less modularity to the container.

\section{Conclusion}

As IoT software embedded on microcontrollers becomes more complex, modularity becomes necessary, and multi-tenant scenarios become more prominent on such hardware. Recent work has focused on enabling the co-location of multiple miniature runtimes and/or virtual machines on a single microcontroller shared by multiple tenants. However, there was so far no adequate  framework available for such multi-tenant scenarios to facilitate finer-grained aspects, including container lifecycle management, adequate container scheduling, and differentiated access control to host RTOS resources. In this paper, we thus introduce TinyContainer, an embedded framework designed for these purposes. We report on benchmarks using our implementation of TinyContainer, which we integrated in a common RTOS to manage small WebAssembly runtimes. We also showcase the advantages of TinyContainer in a TinyML use case whereby access to data and model weights can be isolated from the inference execution engine running in native code, thus benefiting from the best of both worlds: fast execution and isolation for assets that should be protected.


\bibliographystyle{ACM-Reference-Format}
\balance
\bibliography{bibliography,bibliography2,bibliography3}

\end{document}